\documentstyle[prl,aps,preprint,epsfig,amsmath]{revtex}

\newif\ifcolor
\colortrue 

\colorfalse 

\begin{document}
\normalsize
\draft
\widetext

\title{Theory for the optimal control of  time-averaged quantities in 
open quantum systems}
\author{Ilia Grigorenko$^*$, Martin E. Garcia$^{\ddagger}$ and K. H. Bennemann,}
\address{Institut f{\"u}r Theoretische Physik der Freien 
Universit{\"a}t Berlin, Arnimallee 14, 14195 Berlin, Germany,}
\date{\today}
\maketitle

\begin{abstract}
 We present variational theory for  optimal control over a finite
 time interval in  quantum systems with relaxation. The
 corresponding Euler-Lagrange equations determining the optimal control
 field are derived. In our theory the optimal control field fulfills a 
  high order differential equation, which we  solve  analytically for 
 some limiting cases. We determine quantitatively how relaxation 
 effects limit 
 the control of the system.  The theory is applied to  open two level quantum
 systems. An approximate analytical solution for the level 
 occupations in terms of the  applied fields is presented. Different 
other applications are discussed. 
\end{abstract}


The manipulation of quantum mechanical systems by using ultrashort
time-dependent fields represents a challenging fundamental physical
problem.  In the last years, a considerable amount of experimental and
theoretical work was concentrated on designing laser pulses having
optimal amplitude and modulation. Thus the control of the quantum
dynamics in various systems like atoms and
molecules\cite{atomic}, quantum dots\cite{qd}, 
 semiconductors\cite{cole}, superconducting devices\cite{nakamura} and
 Bose-Einstein condensate\cite{cramer} was achieved. 

 Several theoretical studies, most of them using numerical optimization 
techniques,  have shown that it is possible to construct 
 optimal external fields (e.g. laser pulses) to drive a 
 certain physical quantity,  
like the population of a given state,  to reach a 
desired value at a given 
time\cite{drho,Rabitz_path,araujo}. 
 
Although this kind of control might be  relevant for many purposes, a more 
 detailed manipulation of real systems may require the  control of physical 
 quantities over a finite time 
interval. The search for optimal fields able to perform such control  is 
 a much more challenging problem for which no
  theoretical  description has  been given so far. 
 
In this letter we present for the first time an analytical theory for
the control of simple open systems over a finite time
interval. By applying a variational approach we derive a high-order
differential equation from which the optimal control fields are
obtained. We also determine the influence of relaxation, the 
 limits of this control and its
potential applications to the manipulation of fundamental physical
quantities, like the induced current through impurities 
 in semiconductors or the population  of  electronic states at metallic
  surfaces.

Our goal is to formulate a theory which permits to derive explicit 
 equations to be satisfied by the optimal control field. 
 Note that one can guess the form of such equations from general physical 
 arguments. 
  Since memory effects are expected to be important,  
 one should search for a differential equation   containing 
 both the pulse area $\theta(t) = \int_{t_0}^t dt' V(t')$, where
 $V(t)$ is the external field envelope, 
  and its time derivatives. Therefore, for the case 
 of optimal control of dynamical quantities at a given time $t_0$, 
 the  differential equation satisfied by  
 $\theta(t)$ must be of at least second order to fulfill the initial  
conditions $\theta(t_0)$, $\dot{\theta}(t_0)$. In the same way, the 
control of time averaged quantities over a finite time interval $[t_0,t_0 + T]$ with boundary conditions requires a differential equation of at least 
forth order for $\theta(t)$ due to the boundary conditions for $\theta(t)$ and
$\dot{\theta}(t)$ at $t_0$ and $t_0+T$. We show below that for certain open  systems a forth order differential equation  for the control fields arises naturally  using variational approach as an Euler-Lagrange (EL) equation.

We start by considering a quantum-mechanical system 
 which is in contact with the
environment and interacting with an external  
field $E(t)=V(t) \cos(\omega t)$. Here $V(t)$ refers to an arbitrary pulse 
 and $\omega$ is the   carrier frequency.  The evolution of such
system obeys the quantum Liouville equation for the density
matrix $\rho(t)$  with dissipative terms. 
 The control of a time averaged  
 dynamical quantity of the system requires the search for 
 the optimal shape $V(t)$  of the external field. 

 Thus, in order to  obtain the optimal $V(t)$ on time interval $[0,T]$ we propose the following Lagrangian (throughout the paper we use atomic units
$\hbar$=m=e=1)
\begin{eqnarray}
\label{lagrang_main}
L&=&\int_{0}^{T}{A(t)\Big ( \frac{\partial}{\partial t}+
i \hat{\cal{Z}}(t)\Big ) \rho(t) dt} \, + \,  \beta \int_{0}^{T}{ {\cal {L}}_1 dt}.
\end{eqnarray}
 $\beta$ is a Lagrange multiplier and  $A(t)$ is a Lagrange multiplier 
density.
The first term in Eq.~(\ref{lagrang_main}) ensures that the density matrix satisfies the quantum Liouville
equation  with the corresponding Liouville operator 
$\hat{\cal{Z}}(t)$\cite{drho}.  While the first term describes the dynamics of the system under the external field, the  functional ${\cal {L}}_1$ explicitly includes the description of  the optimal control  and is given by 
\begin{eqnarray}
\label{lagrang_second}
 {\cal {L}}_1(\rho,V)={\cal {L}}_{ob}(\rho)+\lambda  {V}^2(t)
+\lambda_1  {\left(\frac{d V (t)}{d t}\right)}^2, 
\end{eqnarray}
where $\lambda$ and $\lambda_1$ are Lagrange multipliers.  ${\cal {L}}_{ob}(\rho)$ refers to a physical quantity to be  
 maximized during the control time. The second term represents a 
constraint on the total energy of the control field 
\begin{eqnarray}
\label{energy}
2 \int_{0}^{T}{E}^2(t)dt\approx\int_{0}^{T}{V}^2(t)dt=E_0.
\end{eqnarray}
The  third term represents a further constraint on the properties of the pulse envelope. The requirement   
\begin{eqnarray}
\label{der}
\int_{0}^{T}{\left(\frac{d V (t)}{d t}\right)}^2 dt \le R,
\end{eqnarray}
 where $R$ is a positive constant, excludes  infinitely narrow or sharp step-like 
 solutions, which cannot be achieved experimentally. 
  
Assuming that the density matrix $\rho(t)$ depends  only on  $\theta(t)$
 and time, one obtains an explicit expression for the functional 
  ${\cal L}_1={\cal L}_1 (\theta,\dot{\theta},\ddot{\theta},t)$. 
 The corresponding  extremum condition $\delta {{\cal L}_1}=0$ yields the
 high-order EL  equation
\begin{equation}
\label{EL_main}
- \lambda_1 \frac{d^4 \theta}{{dt}^4}+ \lambda \frac{d^2 \theta}{{dt}^2} -\frac{1}{2}
\frac{\partial{\cal L}_{ob}(\rho)}{\partial{\theta}}=0. 
\end{equation}
In order to solve Eq.~(\ref{EL_main}) one can assume the natural 
boundary conditions $\theta(0)={\dot \theta}(0)={\dot \theta}(T)=0$, 
$\theta(T)=\theta_T$, which also ensure that $V(0) = V(T) = 0$. 
 The choice of the constant 
$\theta_T$ depends on the problem.  
 In general,  the constants $\theta_T$, $R$ and $E_0$ 
can  be also object of the optimization. 
Note, that above formulated problem is highly nonlinear with respect to 
 the function $\theta(t)$ and can be solved only numerically.

Eq.~(\ref{EL_main}) is the central result of this letter
 and provides  an explicit differential equation for the control field. 
 Note that this equation is only applicable if 
$\rho = \rho (\theta(t),t)$. 
 
In order to show that Eq.~(\ref{EL_main}) can describe optimal control in 
real physical situations, we apply  our theory to an open two level quantum 
 system. This is  characterized by the energy levels   $\epsilon_1$ and $\epsilon_2$, a dipole matrix element $\mu$ and the longitudinal and transverse relaxation constants,  $\gamma_1$ and  $\gamma_2$,   respectively.
 The carrier frequency of the control field is chosen to be the resonant frequency $\omega = \epsilon_2- \epsilon_1$. The dynamics of the density matrix
 $\rho(t)$ follows the equations (in the rotating wave approximation) 
\begin{eqnarray}
\label{Liu}
i \frac{\partial{\rho_{\ell\ell}}}{\partial{t}}
&=&(-1)^{\ell}(\mu V(t)(\rho_{21}-\rho_{12})-i\gamma_1 \rho_{22}),  \nonumber\\
i \frac{\partial{\rho_{12}}}{\partial{t}}
&=&\mu V(t)(\rho_{22}-\rho_{11})-i\gamma_2\rho_{12},\; 
\end{eqnarray}
with $\ell=1,2$. Note that $\rho_{11} + \rho_{22} = 1$ and $\rho_{21}=\rho_{12}^*$. Eqs.~(\ref{Liu}) 
 are used for the  description of different effects, like  for instance, 
 the response of donor impurities in semiconductors to teraherz radiation\cite{cole},  or the excitation of  surface-  into image charge
 states at noble metal surfaces\cite{wolf}. Therefore, the initial conditions 
 are set as  $\rho_{11}=1,\rho_{22} = \rho_{12}=\rho_{21}=0$. 
 
 Eqs.~(\ref{Liu}) have the form $i  \partial \rho(t) / \partial t =
 {\hat{\tilde{\cal{Z}}}}{(t)} \rho{(t)}$ and are difficult to integrate, since 
 $[{\hat{\tilde{\cal{Z}}}}(t),{\hat{\tilde{\cal{Z}}}}(t')] \not= 0$. However,  
 the commutators $[{\hat{\tilde{\cal{Z}}}}(t),{\hat{\tilde{\cal{Z}}}}(t')]$
 become arbitrarily small under the condition\cite{further}
\begin{eqnarray}
\label{condition}
\Big|\frac{\partial\log{ V(t) }}{\partial t}\gamma_{\ell}\Big|\ll1, \; 
\end{eqnarray}
with $\ell=1,2$.
In this case  approximate solution for $\rho_{22}(t)$ is 
\begin{eqnarray}
\label{analyt_form}
\rho_{22}(t)=
2\:\theta^2(t) F^{-1} \Big( 
1-\cosh(H)\exp(-(\gamma_1+\gamma_2)t/2)\nonumber\\
+(\gamma_1+\gamma_2) t\:\sinh(H)
\exp(-(\gamma_1+\gamma_2) t/2) H^{-1}
\Big)
,
\end{eqnarray}
where $H=\sqrt{((\gamma_1-\gamma_2)^2 t^2-16\;\theta^2(t))/2}$, and
$F=\gamma_1 \gamma_2 t^2+4\;\theta^2(t)$. Note that this approximate 
solution  becomes exact  when $\gamma_1=\gamma_2=0$ or for 
a constant control field $V(t)=V_0$.
 The expression of Eq.~(\ref{analyt_form}) has the form $\rho =
 \rho(\theta(t),t)$ and therefore Eq.~(\ref{EL_main}) is applicable.

Now we construct the functional $
{\cal{L}}_{ob} (\rho) =\rho_{22} (t), 
$
 so that the average occupation of the upper level $n_2 = \int_{0}^{T}{\rho_{22} (t) dt}$    is maximized.  
Note, that $n_2$ proportional to the  
observed photocurrent\cite{cole} in teraherz experiments on semiconductors.  
 The resonant tunneling current  through an array of coupled quantum dots 
is also proportional to a such value\cite{nazarov}.

We have calculated the optimal $V(t)$ from the numerical integration of Eq.~(\ref{EL_main}) for different values of the
relaxation constants $\gamma_1$ and $\gamma_2$ and of the energy $E_0$
and the curvature $R$ of the control fields. For simplicity we consider the control interval
$[0,1]$.

In Fig.~1 we show the optimal field
 for an isolated ($\gamma_1 = \gamma_2 = 0$) and 
 for an open two level system for given values of the pulse energy and 
 curvature. 
Note, that for both cases the pulse maximum occurs near the beginning of the control interval. This leads to a rapid increase of the population $\rho_{22}(t)$
 and therefore to a maximization of $n_2$. In the case of an isolated system
 the pulse vanishes when the population inversion has been achieved,  whereas for an open system the pulse must compensate the 
decay of $\rho_{22}(t)$ due to relaxation effects and remains finite over the whole control interval.  
  
In the inset of Fig.~1 we show the corresponding dynamics of the
population $\rho_{22}(t)$ for both cases. As mentioned before, 
Eq.~(\ref{analyt_form}) is exact for the isolated system. 
 Note, that for the open system the analytical form of $\rho_{22}$ 
 (Eq.~(\ref{analyt_form})) compares well with the numerical solution of the Liouville equation. This indicates that $V(t)$ fulfills the
 condition(\ref{condition}) on the control interval.  

 We found that the value of $n_2$ increases both for the isolated and
 for the open system monotonously with energy of the optimal control
 field. In Fig.~2 we plot $n_2$ as a function of the energy $E_0$ and
 the curvature $R$ of the optimal fields obtained from
 Eq.~(\ref{EL_main}). Note, that pulses of fixed shape (for instance
 Gaussian) would show an oscillating behavior for increasing energy
 due to Rabi oscillations \cite{oliver}. The monotonous increase is a feature which
 characterizes the optimal pulses.

In order to achieve a simplified study of 
 the physics contained in the control fields of Fig.~1, we  analyze the problem in certain limiting cases. 
 For instance, if $\gamma_{1,2} T \ll 1 $ one can neglect decoherence within the control interval and Eq.~(\ref{analyt_form}) becomes 
  $\rho_{22}(t)=\sin^2 \left( \theta(t) \right)$.
 In order to make the problem analytically solvable, we reduce the order of the differential equation for the control fields. For that purpose  we replace the  constraint on  the derivative of the field 
envelop  (Eq.~(\ref{der})) by a weaker one obtained from the condition   
\begin{equation}
\label{width_cond}
\int_{0}^{T} {\dot{\theta}}^2(t) dt -\frac{1}{T}{\left( \int_{0}^{T} \dot{\theta}(t) dt\right)}^2  = \int_{0}^{T} {\cal{L}}_w(\theta) 
\ge S,
\end{equation}
where S is a positive constant. Eq.~(\ref{width_cond})  merely 
 bounds the width of the envelope $V(t)$ in 
 order to avoid unphysically narrow pulses.   
  Thus, the Lagrangian density ${\cal{L}}_1$ for the optimal control has the form 
\begin{equation}
\label{lagrang_simple}
{\cal{L}}_1 = {\rho_{22}(t) } +\lambda  \dot{\theta}^2(t) +\lambda_2 {\cal{L}}_w(\theta),
\end{equation}
while the corresponding EL equation is given by 
\begin{equation}
\label{sine-gordon}
2\lambda' \ddot{\theta}(t)-\sin(2\:\theta(t))=0.  
\end{equation}
 Note, that condition (\ref{width_cond}) only leads to a 
rescaling of the Lagrangian multiplier $\lambda$ to $\lambda'=\lambda+ \lambda_2$. 
 The second order differential Eq.~(\ref{sine-gordon}) requires two boundary conditions, for which we choose 
${\theta}(0)=0$ and ${\theta}(T)=\pi/2$ 
(which ensure the population inversion).
 Eq.~(\ref{sine-gordon}) resembles that for a mathematical pendulum  and 
 can be solved analytically. The resulting field envelope is given by 
\begin{equation}
\label{solution}
V(t) = V(0) \, dn(V(0)t,C),
\end{equation}
where $dn$ is the Jacobian elliptic function, and $C=-{(\lambda' V^2(0))}^{-1}$ is a constant of integration.  Note, that $V(0) \not= 0$. 
 Using conditions (\ref{energy}) and (\ref{width_cond})
 we determine coefficients $\lambda$ and $\lambda_2$. 
If we choose $C\to 1$ then we can obtain $V(T)\to 0$. In this case, Eq.~(\ref{solution}) can be significantly simplified to $
V(t) = \frac{\partial}{\partial t} 
{ 
 \arccos
    [2 \exp{(V(0) t)} / 
    ( 1+\exp{(2 V(0) t)} ) ]
}. 
$

 In Fig.~3 we plot the optimal control field $V(t)$ which maximizes 
 the Lagrangian (\ref{lagrang_simple}) for isolated and open two level systems. In both cases the field has its maximum value at $t = 0$ and exhibits a monotonous decay. As in the case of the solutions of the forth-order Eq.~(\ref{EL_main}) the control 
 field is broader for the open system.  In the inset of Fig.~3 we plot the population $\rho_{22}(t)$. The overall behavior of $\rho_{22}(t)$ 
is similar to that of the populations shown in Fig.~1. 

 It is important to point out, that  a Lagrangian of the form of  Eq.~(\ref{lagrang_simple}) always leads to a second order differential equation for the control fields as long   as the condition $\rho = \rho(\theta(t),t)$ is satisfied.
 Therefore,   one cannot demand extra boundary   
  conditions for the fields  $V(0)=V(T)=0$. Otherwise one would 
obtain the trivial solution $V(t)\equiv 0$, which 
 is not consistent with either (\ref{energy}) or (\ref{width_cond}). 
Therefore, if conditions on $V(0)$ and $V(T)$ have to be imposed, 
 a Lagrangian leading to a forth order differential equation is 
necessary, as we have shown before.

As it was mentioned before $n_2$ increases monotonously with the pulse energy for the optimal field. Since for the isolated system $n_2$ approaches the maximum possible value $n_2=T$, in the case of nonisolated systems there is a limit. 
 In order to show that this limits is due to general physical reasons we 
analyze  the occupation $\rho_{22}(t)$ (Eq.~(\ref{analyt_form})) in more detail. 
 For a strong control field satisfying  
$\gamma_{1,2} t /  \theta(t)  \ll 1 $ 
 the occupation $\rho_{22}(t)$  always lies under the curve  $\rho_{22}^{max}(t)=(1+\exp(-(\gamma_1+\gamma_2) t/2))/2$. This means that it 
 exhibits an absolute upper bound. 
 Therefore due to dissipative processes the following inequality holds for the controlled averaged value of $\rho_{22}$:  
\begin{eqnarray}
\label{limit}
n_2 = \int_{0}^{T}{\rho_{22} (t) dt} \leq T/2+(1-\exp(-(\gamma_1+\gamma_2)T/2)/( \gamma_1+\gamma_2).
\end{eqnarray}

Eq.~(\ref{limit}) shows the absolute limit for the optimal 
control of averaged occupations 
 in  open two level systems. In Fig.~4 we show the maximal
 possible value $\rho_{22}^{max}(t)$  and the time evolution of $\rho_{22}(t)$ induced by 40 randomly generated pulses (for some of  which the condition $\gamma_{1,2} t /  \theta(t)  \ll 1 $  is even not strictly fulfilled). 
 From Fig.~4 we conclude that under the action of arbitrary control 
fields, the life time of the upper level cannot be longer than $2/(\gamma_1 + \gamma_2)$.

Using this result we can  determine the maximal possible life-time 
 for an image state at a  
Cu(111) surface which can be achieved by pulse shaping. According to 
 Hertel et al.\cite{wolf}, those states are characterized by $\gamma_1 = 5 \cdot 10^{13} s^{-1}$ and $\gamma_2 = \gamma_1/2$.
 Thus, our theory predicts in that case an effective  decay 
constant $\gamma_{eff}=(\gamma_1+\gamma_2)/2 = 3.75 \cdot 10^{13} s^{-1}$.

In summary, we presented a theory for the description of 
 optimal control of time-averaged quantities 
in open quantum systems.  
 In particular we have shown that the boundary conditions of the problem 
 make a significant influence on the shape of the optimal fields. 
  In contrast to other approaches our theory allows to derive
an explicit differential equation for the optimal control field,
 which we integrated both numerically and exactly for some limiting cases. 
Our approximation $\rho(t)=\rho(\theta,t)$
 was checked by direct integration of the Liouville
 equations and it seems to hold also in the case of strong relaxation. 
Using our theory we found the optimal fields which maximize the 
 population  of the upper levels of isolated and open two-level systems.
 We found an absolute upper bound for this kind of optimal control.
 Our approach can be used for further investigations, for instance,control
 of the dynamics of multi-level systems.

\begin{figure}[h]
\begin{center}
\includegraphics[height=\textwidth,angle=-90,bb = 100 110 600 750 ]{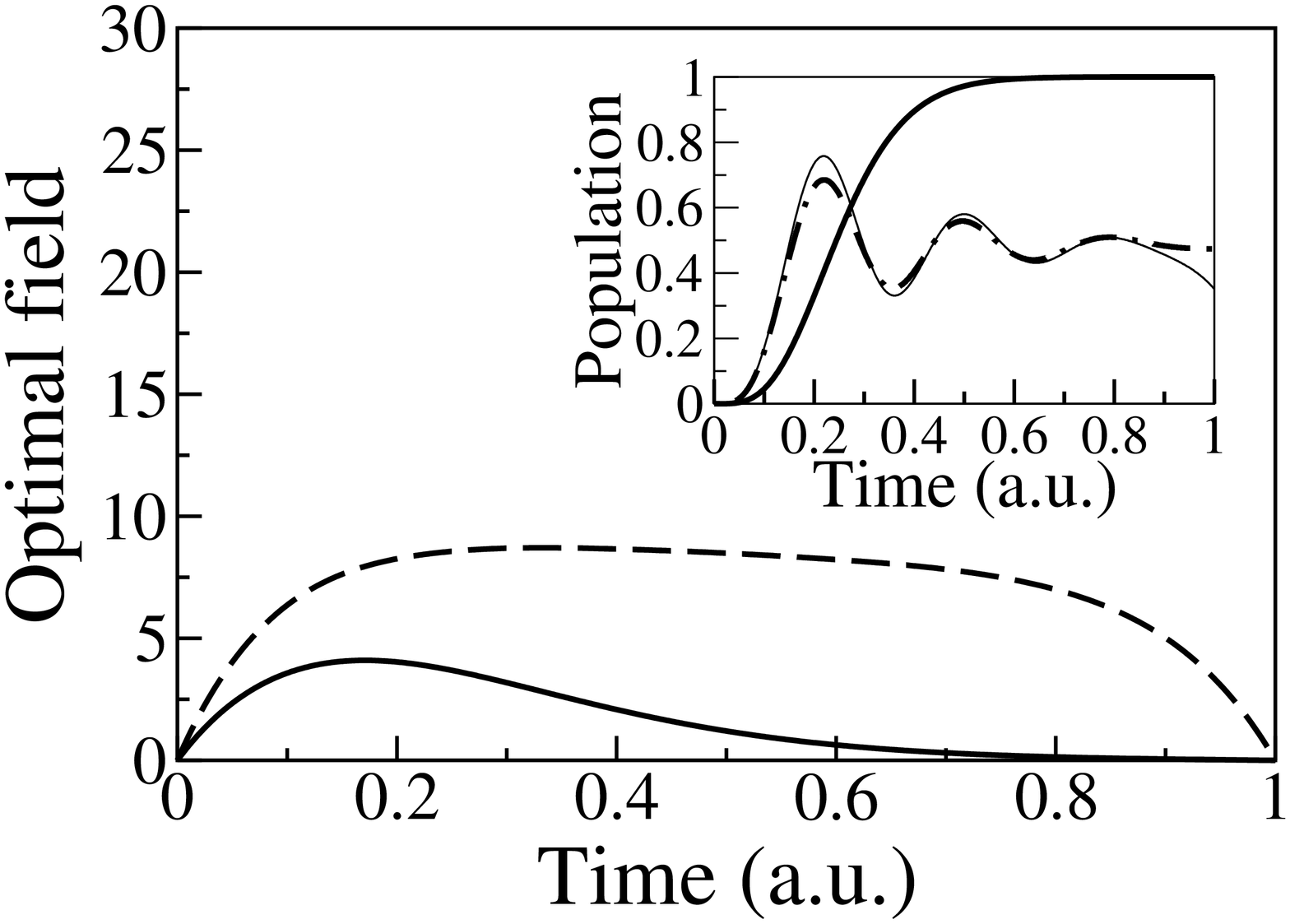}
\caption { Optimal control field for an isolated two level system  ($\gamma_{1,2}=0$, solid line). The pulse energy is $E_0=4.57$ and the pulse 
curvature $R=128.4$. The dashed line shows the optimal pulse for 
 the open system ($\gamma_{1}=2 \gamma_{2}=5$) with 
 energy $E_0=53.54$ and  curvature $R=808.8$.  
   Inset: Dynamics of the occupation $\rho_{22}(t)$ for 
 an isolated system (thick solid line) and with relaxation 
(dash dotted line-using formula (\ref{analyt_form}), thin solid line-numerical solution of the Liouville equation (\ref{Liu})). Arbitrary units are used.}
\end{center}
\end{figure}
\begin{figure}[h]
\includegraphics[height=0.7\textwidth]{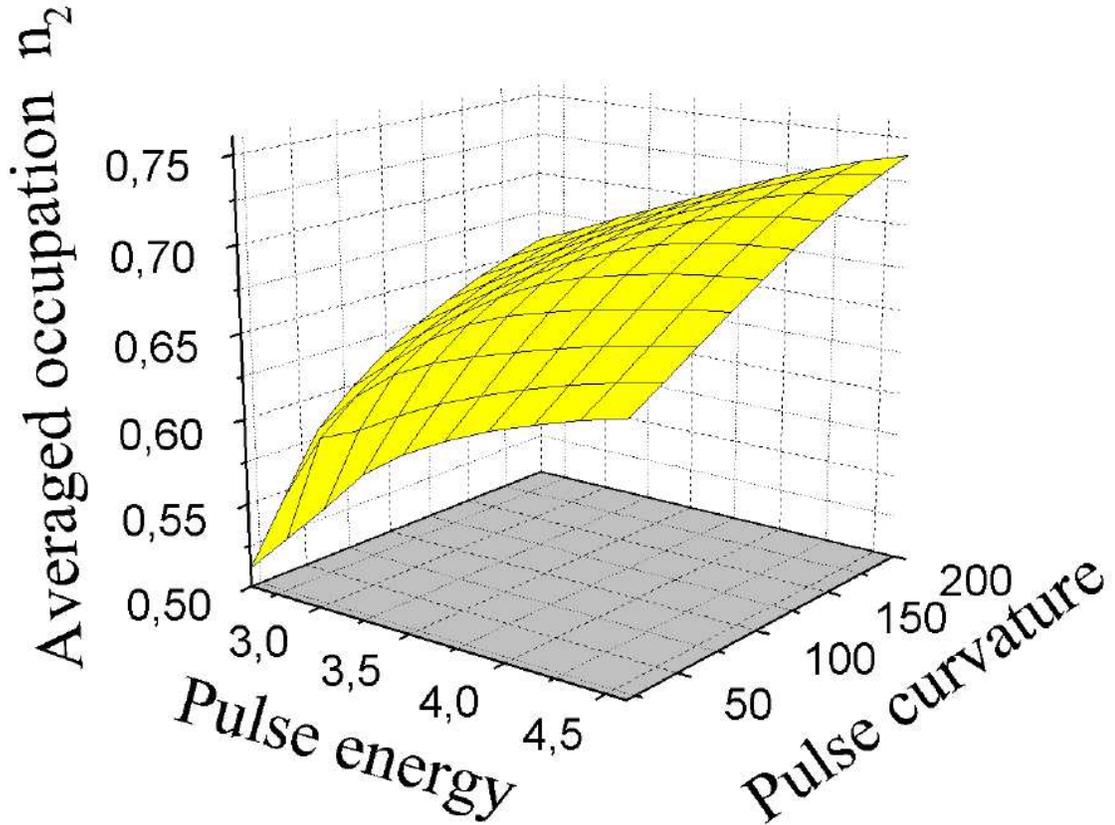}
\caption {Dependence of the averaged occupation $n_2$ as a function of the energy $E_0$ and the curvature $R$ (see Eq.~(\ref{der}) ) of the optimal pulses. }
\end{figure}
\begin{figure}[h]
\begin{center}
\includegraphics[height=\textwidth,angle=-90]{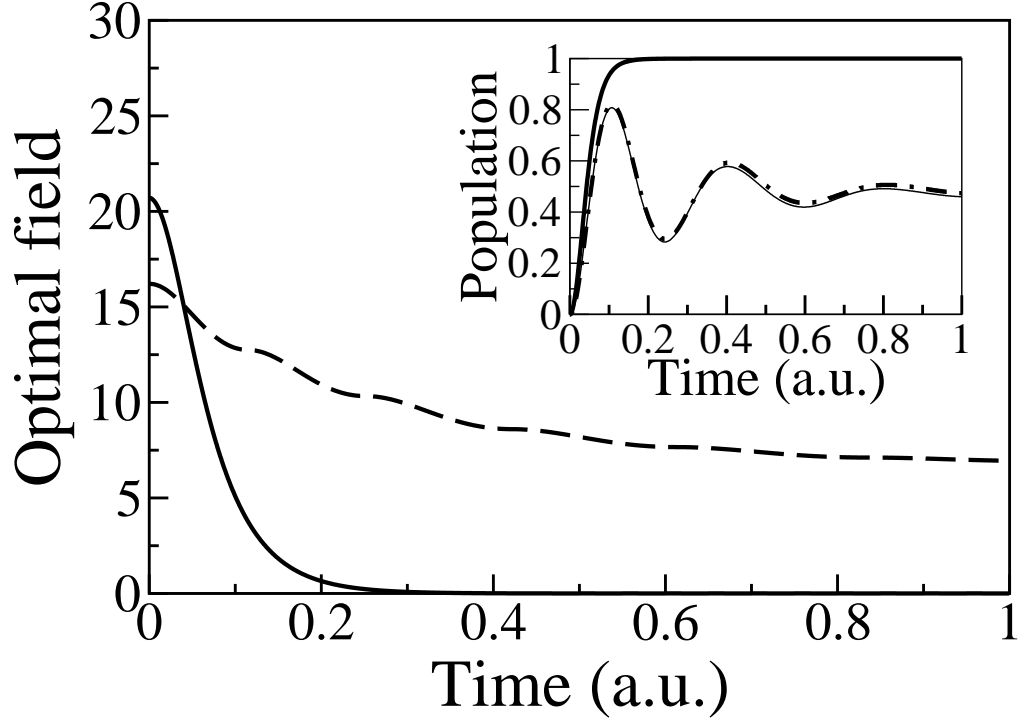}
\caption {Optimal control field for an isolated two level system ($\gamma_{1,2}=0$, solid line). The pulse energy is $E_0=20.50$. Dashed line: optimal field 
 for the open system  ($\gamma_{1}=2 \gamma_{2}=5$) with a pulse energy  $E_0=89.72$. 
Inset: Dynamics of the occupation $\rho_{22}(t)$ for an isolated system (thick solid line) and with relaxation 
(dash dotted line using Eq.~(\ref{analyt_form}), thin solid line-numerical solution of the Liouville equation (\ref{Liu})). }
\end{center}
\end{figure}
\begin{figure}[h]
\begin{center}
\includegraphics[height=0.8\textwidth,angle=-90,bb = 100 110 600 750 ]{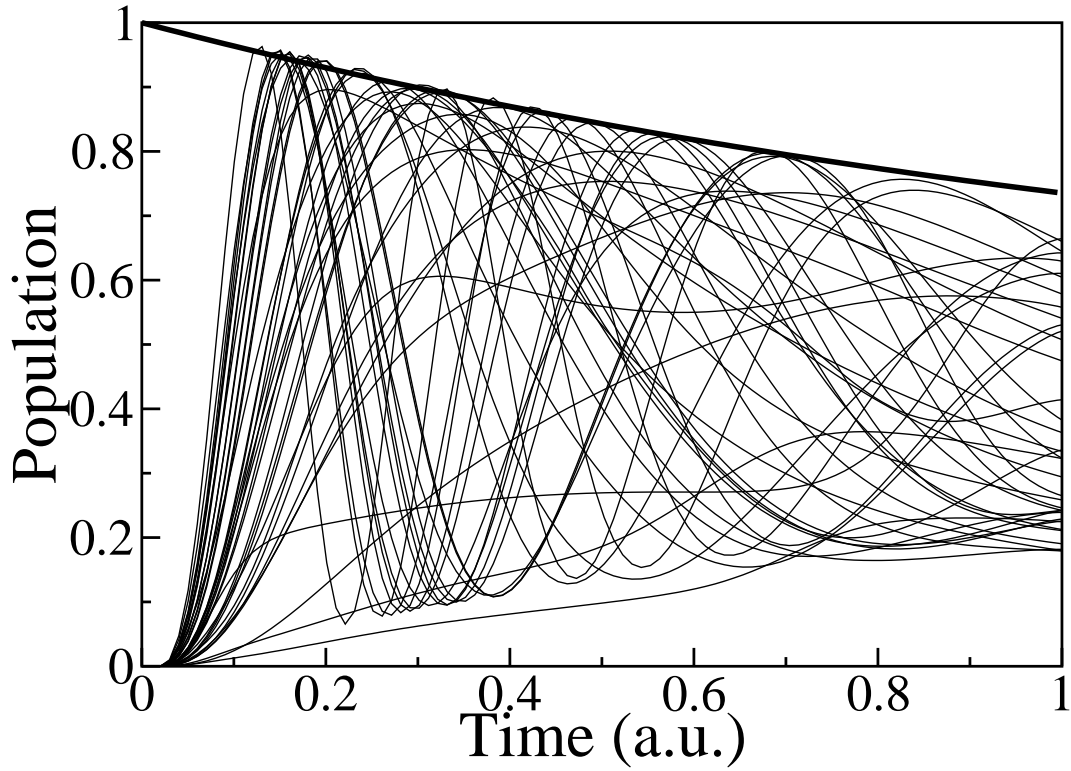}
\caption {Dynamics of the occupation $\rho_{22}(t)$  for 40 
randomly generated control pulses and for $\gamma_{1}=2\gamma_{2}=1$ 
(thin solid lines). The thick solid line represents a bound for the possible values of $\rho_{22}^{max}(t)=(1+\exp(-(\gamma_{1}+\gamma_{2}) t/2))/2$. }
\end{center}
\end{figure}


\begin{thebibliography}{99}
\bibitem{further} further details will be published elswhere.
\end{thebibliography}
\end{document}